\documentstyle[12pt]{article}
\def\z{\zeta}
\def\zb{\bar{\zeta}}
\def\beq{\begin{equation}}
\def\eeq{\end{equation}}
\def\bea{\begin{eqnarray}}
\def\eea{\end{eqnarray}}
\def\MPL{{\it Mod. Phys. Lett.} }
\def\NP{{\it Nucl. Phys.} }
\def\PL{{\it Phys. Lett.} }
\def\PR{{\it Phys. Rev.} }
\def\ap{\alpha ^{\prime}}
\begin {document}
\begin{titlepage}
April 1998 \\
\begin{flushright}
HU Berlin-EP-98/21\\
\end{flushright}
\mbox{ }  \hfill hepth@xxx/9804065
\vspace{5ex}
\Large
\begin {center}
\bf{The mass term in non-Abelian gauge field dynamics on matrix D-branes
and T-duality in the $\sigma$-model approach }
\end {center}
\large
\vspace{1ex}
\begin{center}
H. Dorn
\footnote{e-mail: dorn@qft3.physik.hu-berlin.de}
\end{center}
\normalsize
\it
\vspace{1ex}
\begin{center}
Humboldt--Universit\"at zu Berlin \\
Institut f\"ur Physik, Theorie der Elementarteilchen \\
Invalidenstra\ss e 110, D-10115 Berlin
\end{center}
\vspace{4ex}
\rm
\begin{center}
{\bf Abstract}
\end{center}
The formal extension of the T-duality rules for open strings from
Abelian to non-Abelian gauge field background leads in a well known
manner to the notion of matrix valued D-brane position. The application
of this concept to the non-Abelian gauge field RG $\beta $-function
of the corresponding $\sigma $-model yields a mass term in the
gauge field dynamics on the matrix D-brane. 
The direct calculation in a corresponding
D-brane model does $not$ yield such a mass term, if the Dirichlet
boundary condition is implemented as a constraint on the integrand in
the defining functional integral. However, the mass term arises in the direct
calculation  for a D-brane model with dynamically realized boundary condition.
\vfill      
\end{titlepage}
\setcounter{page}{1}
\pagestyle{plain}
\section{Introduction}
T-duality for open strings interchanges Neumann and Dirichlet boundary
conditions \cite{dual,polrev}. If in a more general setting the string couples with its 
free ends to an Abelian gauge field, the boundary condition stating the
balance of the normal derivative of the world sheet embedding and the Lorentz force
$$\partial _{\bf n}X^{\mu}~+~2\pi\ap F^{\mu}_{~~\nu}\partial _{\bf t}X^{\nu}
~=~0$$
turns under the T-duality for the $i^{\mbox{\footnotesize th}}$ coordinate 
$\partial _{\bf n}X^i\rightarrow\partial _{\bf t}X^i$ ({\bf n} and {\bf t}
denote normal and tangential, respectively) into the condition
$$X^i~=~2\pi\ap A^i(X^M)~,$$
if the gauge field $A_{\mu}$ depends only on a subset $X^M$ of the coordinates
$X^{\mu}=(X^i,X^M)$. Thus in the dual description the string endpoints have to
move on the manifold, the D-brane, defined by the gauge fields in the $i$-directions.

The formal extension of this recipe to non-Abelian gauge fields \cite{witten}
leads to the notion of matrix valued D-brane positions which plays
a crucial role in the heuristic motivation of M(atrix) theory \cite{banks} and
the emergence of non-commutative geometric structures in string theory \cite{doug}.

In this paper we will comment on the issue of a $\sigma $-model description
of the T-dual model in the case of non-Abelian gauge fields including
its quantum corrections and the issue of realizing the boundary condition
either as an external one or a dynamical one.
By external realization we mean a condition on the fields which are 
integrated over in
the functional integral, by dynamical realization we have in mind conditions
arising as part of the stationarity condition of the action. Both kinds
of realization are equivalent on the classical level, but may lead to different
quantum theories.

The formal manipulations of the functional integral for the $\sigma $-model, 
describing the motion of a closed string in nontrivial target space fields
$(G,~B,~\Phi )$ related to its massless excitations, which yield Buscher's
duality rules \cite{bu}, have been extended to a $\sigma $-model on a manifold with boundary which in addition couples to a gauge field in ref.\cite{do},
see also \cite{alv}. There a dual model with externally realized Dirichlet
boundary conditions is generated. The RG $\beta $-functions of this model
have been calculated in ref.\cite{leigh} already. The resulting conformal 
invariance conditions are equivalent to the stationarity conditions of
the Born-Infeld action dimensionally reduced to the D-brane \cite{leigh}.
The Born-Infeld action is, up to the dimensional reduction, form invariant
under the T-duality transformation \cite{alv}.
Therefore, at least up to the implemented level of accuracy, the gauge field
part of the naive T-duality rule is compatible with renormalisation.
\footnote{See also the discussion in the second ref. of \cite{do} and the
discussion of T-duality and renormalisation for the closed string fields
\cite{haage}.}

With the help of an one-dimensional auxiliary $\z $-field formalism 
\cite{zform,gn}
resolving the path ordering prescription for the Wilson loop we were able
in \cite{do,d} to extend the treatment to the case of non-Abelian gauge
fields. In this way the notion of matrix valued brane position got
a special realization suitable for practical calculations. Postponing the
$\z $-integration until the very end, at intermediate stages the non-Abelian
gauge field appears sandwiched between $\zb $ and $\z $ only
$$\zb _a(s) (A_{\mu}(X(z(s)))_{ab}\z _b(s)~.$$
Hence the only difference to the Abelian situation is an explicit dependence 
on the parameter $s$, parametrising the boundary of the string world sheet.
The gauge field and brane position RG $\beta $-functions have been calculated
in lowest order of $\ap $ in \cite{d}.

Our only focus in this paper concerns the mass term for non-Abelian gauge field
dynamics on matrix D-branes which arises by direct application of the
naive T-duality rule to the $\beta $-function of the dual partner, i.e.
the usual open string with free ends moving in target space non-Abelian
gauge field background. Such a term does {\it not} appear in the direct calculation of the D-brane model \cite{d,dproc}. Thus at this state of affairs the 
non-Abelian gauge field 
part of the naive T-duality transformation seems to be incompatible with
renormalisation at lowest order, already. Before sketching the plan of the
paper we will recall the formal duality argument giving rise to the
mass term \cite{dproc}.

The lowest order gauge field $\beta $-function for the string with free ends
is
\footnote{We denote all quantities in this model with a tilde. The untilded
quantities refer to the D-brane model.}\cite{zeit,brech}
\beq
\tilde{\beta}^{(\tilde A)}_{\lambda}~=~-\ap ~\eta ^{\mu\nu}~\tilde{D}
_{\mu}\tilde{F}_{\nu\lambda}~.
\label{1}
\eeq
If e.g. $\partial _1 A_{\mu}=0$ we get, using 
$X^1=f(X^M)=-2\pi\ap \tilde A_1(\tilde X^M),~\eta ^{11}=-1$\\
and $\tilde X^M=X^M,~A_1=0,~ A_M=\tilde A_M$,
\bea
\tilde{\beta}^{(\tilde A)}_L&=&-\ap ~\eta ^{MN}~D_MF_{NL}~+~\frac{i}
{4\pi ^2\ap}\left [f,D_Lf\right ]~,\nonumber\\
\tilde{\beta}^{(\tilde A)}_1&=&\frac{1}{2\pi}\eta ^{MN}~D_MD_Nf~.
\label{2}
\eea
The second term in $\tilde{\beta}^{(\tilde A)}_{\lambda}$ turns out as a
standard mass term, at least in the special D-brane
configuration of diagonal and constant $f$, i.e. $f_{ab}=f_a\delta _{ab},~
\partial _Mf_a=0$, which describes planar parallel D-brane copies at position
$X^1=f_a,~a=1,...,n$. The equation of motion obtained from 
$\tilde{\beta}^{(A)}_L=0$ then is  
\beq
(D^M~F_{ML})_{ab}+\left (\frac{1}{2\pi\ap}\right )^2 (f_a-f_b)^2 (A_L) _{ab}~=~0~.
\label{3}
\eeq
As expected the mass is proportional to the separation of the D-brane copies.

The bulk of the paper is organised as follows. To keep things as clear as
possible we restrict ourselves to the case of trivial closed string background
fields and consider T-duality in one direction ($X^1$) only. The calculations
crucially depend on the $\z $ auxiliary formalism. To become familiar with
this calculus we start in section 2 with the derivation of the well known
variational formulae for the standard Wilson loop \cite{wloop}. We do this in a 
rather explicit way to amplify the step crucially also in the later modified
applications: The use of the $\zb $, $\z $ equation of motion in correlation
functions including contact terms.

To present a simple check of our results in \cite{d,dproc} 
free of the subtleties connected with the covariant expansion of the imposed
boundary constraint, we use in section 3 the simplification due to flat
target space to integrate the functional $\delta $-function. Things become 
even simpler if one then considers constant $f$ only. But also in this
very transparent situation no mass term appears.

In section 4 we switch to dynamically realized Dirichlet conditions. The
corresponding model is formulated quantising the action obtained
in our paper \cite{do} by a canonical transformation of the free end model.
As it will turn out, this model yields a gauge field $\beta $-function with
mass term. In the final section we summarise and interprete our result.
In particular it will be stressed that the reason for the 
discrepancy of both versions is strongly related to a partial interchange
of the functional integration over the string position with that over
the auxiliary $\z $-field.
  
\section{Derivation of the standard variational formula for the Wilson loop
in $\z$-language}
The Wilson loop as a functional of a closed path $X^{\mu}(s),~0\leq s\leq 1,~
X^{\mu}(0)=X^{\mu}(1)$ defined by
\beq
{\cal W}[X] ~=~\mbox{tr P}\exp \left (i\int ^1_0 A_{\mu}(X(s))\dot X^{\mu}~ds
\right )
\label{4}
\eeq
can be expressed by the following functional integral over an one-dimensional 
auxiliary field $\z ,~\zb$ with propagator $\langle \zb _a(t)\z _b(s)\rangle =
\delta _{ab} \Theta (s-t)$ \cite{zform,gn}
\beq
{\cal W}[X] ~=~\int D\bar{\z}D\z ~\zb _a(0)\z _a(1)~e^{iS_0[\z ,\zb]}~W[X,\z ,\zb ]~.\label{5} 
\eeq
$S_0$ and $W$ are defined by
\beq
S_0[\z ,\zb]~=~i\int_0^1\zb (s)\dot{\z}(s)ds~+~i\zb (0)\z (0)\label{6}
\eeq
and
\beq
W[X,\z ,\zb ]~=~\exp \left (i\int _0^1\zb A_{\mu}(X(s))\z (s)~\dot X ^{\mu}ds
\right )~.
\label{7}
\eeq
Perhaps it is useful to stress that due to the sum over the index $a$,
realizing the trace, there is no dependence on the choice of the point
on the closed path from which we count $s$.
Varying with respect to $X$ and performing one partial integration we get
$$
\frac{\delta W}{\delta X^{\mu}(t)}~=~iW\left (\zb (\partial _{\mu}A_{\lambda}
-\partial _{\lambda}A_{\mu})\z \dot X^{\lambda}(t)-\dot{\zb}A_{\mu}\z-\zb A_{\mu}
\dot{\z}    \right )~.
$$
Now the use of the equation of motion for $\z $ and $\zb $ just produces
the commutator term, necessary to complete the full non-Abelian field 
strength. To verify this we note that for the full $\z $-action
\beq
S[\z ,\zb ]~=~S_0[\z ,\zb]-i\log W~=~S_0[\z ,\zb]+\int _0^1\zb A_{\mu}\z \dot X^{\mu}ds\label{8}
\eeq
one has (the corresponding eq. for $\dot{\zb}$ is obvious)
\beq
\dot{\z}~=~iA_{\mu}\z\dot X^{\mu}-i \frac{\delta S}{\delta \zb (s)}~.
\label{9}
\eeq
The argument is then completed by the application of
\beq
\int D\zb D\z ~\zb _a(0)\z _a(1)~e^{iS[\z , \zb]}~h(\z (t),\zb (t))~\frac{\delta S}
{\delta \zb (t)}~=~0~,\label{10}
\eeq
which is valid for an arbitrary polynomial $h$. Altogether we arrive at
\footnote{Here and in the following we understand all equations for $W$
and its derivatives up to terms vanishing under the $\z ,\zb $-integration.} 
\beq
\frac{\delta W}{\delta X^{\mu}(t)}~=~iW~\zb F_{\mu\lambda}\z ~\dot X ^{\lambda}~,
\label{11}
\eeq
i.e.
\beq
\frac{\delta{\cal W}}{\delta X^{\mu}(t)}~=~i~\mbox{tr P}\left (e^{i\int _0^1
A_{\nu}\dot X^{\nu}ds}F_{\mu\lambda}\dot X^{\lambda}(t)\right )~.\label{12}
\eeq
The second functional derivative becomes
\bea
\frac{\delta ^2~W}{\delta X^{\mu}(t)\delta X^{\nu}(t')}&=&
iW~\delta (t-t')~\zb \partial _{\nu}F_{\mu\lambda}\z ~\dot X^{\lambda}(t)
~+~iW~\dot{\delta}(t-t')~\zb F_{\mu\nu}\z\label{13}\\
&-&W~\zb F_{\mu\lambda}\z ~\dot X^{\lambda}(t)~\left ( \zb (\partial _{\nu}A_{\kappa}-\partial _{\kappa}
A_{\nu})\z ~\dot X^{\kappa}(t')-\dot{\zb}A_{\nu}\z -\zb A_{\nu}\dot{\z}(t')\right )~.\nonumber 
\eea
In the bilocal $(t,t')$-term we again want to use the $\z $-equations of motion
to get the Abelian part of the field strength at $t'$ completed by the commutator term to the full non-Abelian field strength. But now there is a second
factor at parameter value $t$ which is responsible for the appearance of a 
contact term $\propto \delta (t-t')$ which just covariantizes the derivative
of $F_{\mu\lambda}$ in the first term of the r.h.s. of (\ref{13}). The underlying general formula is
($k$ and $h$ polynomials in $\z $, $\zb $)
\bea
\lefteqn{\int D\zb D\z ~\zb _a(0)\z _a(1)~e^{iS[\z , \zb]}} &&\\
&\cdot&\left (ih(\z (t),\zb (t))~\frac{\delta S}{\delta \zb (t)}~k(\z (t'),\zb (t'))
~+~ \delta (t-t')~h((\z (t),\zb (t))~\frac{\partial k}{\partial \zb (t)}\right )
~=~0~.\nonumber
\label{14}
\eea
Applying this to eq.(\ref{13}) we get
\bea
\frac{\delta ^2~W}{\delta X^{\mu}(t)\delta X^{\nu}(t')}&=&
iW~\delta (t-t')~\zb D _{\nu}F_{\mu\lambda}\z ~\dot X^{\lambda}(t)
~+~iW~\dot{\delta}(t-t')~\zb F_{\mu\nu}\z\label{15}\\
&-&W~\zb F_{\mu\lambda}\z ~\dot X^{\lambda}(t)~\zb F_{\nu\kappa}\z ~\dot X^{\kappa}
(t')~.\nonumber 
\eea
The transition to $\frac{\delta ^2~{\cal W}}{\delta X^{\mu}(t)\delta X^{\nu}(t')}$ is, similar to the pair (\ref{11}),(\ref{12}), obvious and will not be written down. Finally we note that (\ref{11}) and (\ref{15}) can be
summarised in
\bea
\lefteqn{W[X+Y ,\z ,\zb ]~=~W[X,\z ,\zb ]}&&\\
&&\cdot ~\exp \left (i\int _0^1\zb (F_{\mu\lambda}
\dot X^{\lambda}Y ^{\mu}+\frac{1}{2}D_{\mu}F_{\nu\lambda}\dot X^{\lambda}
Y ^{\mu}Y ^{\nu}+\frac{1}{2}F_{\mu\nu}Y ^{\mu}
\dot Y^{\nu})\z ds+
O(Y^3)\right )~.\nonumber 
\label{16}
\eea
\section{$\beta $-functions in the case of external Dirichlet boundary 
condition}
We consider the simplest case of only one coordinate of type $i$ and call 
this $X^1$. Then the D-brane is defined by $X^1~=~f(X)$ with 
$\partial _1f=0$. 
$A_M(X)$, $\partial _1 A_M(X)=0$ describes the gauge field on the brane. Both
$A$ and $f$ are matrix-valued. The string world sheet $M$ is parametrized by 
the two-dimensional coordinate $z$, the boundary $\partial M$
in $z$-space is described by $z(s)$. The partition function for the model we will 
discuss in this section is given by
\bea
\lefteqn{\hat{{\cal Z}}[A,f]~=~\int DX^{\mu}D\zb D\z ~\zb _a(0)\z _a(1)~e^{iS_0[\z , \zb]}
~\delta _{\partial M}(X^1-\zb f(X)\z)}&&\nonumber\\
&&\cdot \exp \left (\frac{i}{4\pi\ap}\int _Md^2z(\partial
X^N\partial X_N+\partial X^1\partial X_1)+i\int _{\partial M}\zb A_N\z ~\dot X^Nds
\right )~.
\label{17}
\eea
It is a special case of the model discussed in ref.\cite{d}. In a first step
we can perform the $X^1$-integration
\bea
I[\zb f\z ]&\equiv &\int DX^1 \exp \left (\frac{i}{4\pi\ap }\int _Md^2z~\partial
X^1\partial X_1\right )\delta _{\partial M}(X^1-\zb f(X)\z)\nonumber\\
&=&(\mbox{det}\partial ^2)^{-\frac{1}{2}}\exp\left (\frac{i}{4\pi\ap}\int _{\partial M}\bar X^1\partial _{\bf n}\bar X_1ds\right )~.
\label{19}
\eea
The functional determinant of the Laplacian is taken within the space of functions approaching the value zero at the boundary.
In the above equation $\bar X^1$ is the unique solution 
\footnote{We switch to Euclidean two-dimensional world volume.}
of the boundary value problem
\beq
\partial ^2\bar X^1~=~0~~\mbox{in}~~M~,~~~~~~~~\bar X^1~=~\zb f\z ~~\mbox{on}~
\partial M~.
\label{20}
\eeq
Up to now there was no need to specify any periodicity condition for $\z ,\zb $. However, since the boundary value problem is well posed for $\zb f\z (1)=
\zb f\z (0)$ only, we now choose $\z $ and $\zb $ anti-periodic \cite{gn}.
Then the propagator in the fundamental interval $0<s<1$ is still given by
$\delta _{ab}\Theta (s-t)$. The periodic choice for $\z ,\zb $ would
lead to an saw-toothed propagator which no longer could organise the path ordering.

The solution of (\ref{20}) can be represented with some kernel $p(z,s)$ fixed by the geometry of $M$
as (For a circle the corresponding integral is the familiar Poisson integral.)
\beq
\bar X^1(z)~=~\int _0^1ds ~p(z,s)\zb (s)f(X^M(z(s)))\z (s)~.
\label{21}
\eeq
Denoting the boundary value of $\partial_{\bf n}p(z,t)$ by
$q(s,t)$ we get (skipping the irrelevant determinant factor)
\beq
I[\zb f\z ]~=~\exp\left (\frac{-i}{4\pi\ap }\int _{\partial M}\int _{\partial M}\zb f\z (s)~
\zb f\z (t)~q(s,t)dsdt\right )~.
\label{21a}
\eeq
Insertion of this result into (\ref{17}) implies
\bea 
\hat {\cal Z}[A,f]&=&\int DX^M D\zb D\z ~\zb _a(0)\z _a(1)~e^{iS_0[\z , \zb]}
\nonumber\\
&&\cdot ~\exp \left (\frac{i}{4\pi\ap}\int _Md^2z~\partial X^N\partial X_N\right )\cdot \hat W[X,\z ,\zb ]~,
\label{22}
\eea
where we introduced
\beq
\hat W[X,\z ,\zb ]~=~\exp\left (i\int _{\partial M}\zb A_N\z \dot X^Nds
~-~\frac{i}{4\pi\ap }\int _{\partial M}\int _{\partial M}\zb f\z (s)
\zb f\z (t)q(s,t)dsdt\right )~.
\label{23}
\eeq
It is very crucial that the modified Wilson loop $\hat W[X,\z ,\zb ]$ is 
a functional of $X^N(z(s))$ only.

Up to now $f(X)$ was an arbitrary matrix-valued function of the coordinates 
$X^N$. To keep the variational formulae for $\hat W[X,\z ,\zb ]$ as simple
as possible, we restrict ourselves in the rest of this section to $constant$ $f$.
Then the bilocal term in $\hat W[X,\z ,\zb ]$ has influence on the $\z ,\zb $ equations
 of motion only. The total action for $\z ,\zb $ instead of $S$ from section 2 is 
\bea
\hat S[\z ,\zb ]&=&S_0[\z ,\zb]-i\log \hat W\\
&=&S_0[\z ,\zb]+\int _0^1\zb A_{N}\z \dot X^Nds-\frac{1}{4\pi\ap }\int _{\partial M}\int _{\partial M}\zb f\z (s)~
\zb f\z (t)~q(s,t)dsdt~,\nonumber
\label{24}
\eea
leading to
\beq
\dot{\z} (t)~=~iA_{N}\z \dot X^{N}(t)-\frac{i}{2\pi\ap}f\z (t)\int _0^1
\zb f\z (s)q(t,s)ds~-~i \frac{\delta \hat S}{\delta \zb (s)}~.
\label{25}
\eeq
Note that the integral in the last line just represents $\partial _{\bf n}X^1$. Therefore the equation of motion obtained by setting $\frac{\delta \hat S}{\delta \zb (s)}$ to zero coincides with the formal dualization of 
(\ref{9}). Repeating now with the modified quantities the steps presented in section 2 we get
\beq
\hat W[X+Y ,\z ,\zb ]~=~\hat W[X,\z ,\zb ]~\exp(i\hat{\Omega} [X,Y,\z ,\zb ]
)~,\label{26}
\eeq
with
\bea
\lefteqn{\hat{\Omega}[X,Y,\z ,\zb ]~=~O(Y^3)}&&\nonumber\\
&&+\int _0^1\bigg \{\bigg ( \zb F_{ML}\z (s)
\dot X^{L}-\frac{i}{2\pi\ap }\zb [f,A_M]\z (s)\int _0^1
\zb f\z (t)~q(s,t)dt\bigg )Y^M \nonumber\\
&&~~~~~~~~~~~~~~~~~~~+\frac{1}{2}\zb D_MF_{NK}\z (s)~\dot X^KY^MY^N~+~\frac{1}{2}\zb F_{MN}\z (s)~Y ^M\dot Y^{N}\nonumber \\
&&~~~~~~~~~~~~~~~~~~~~~~~~-\frac{i}{4\pi\ap }\zb D_M[f,A_N]\z (s)~Y^M Y^N\int _0^1
\zb f\z (t)q(s,t)dt\bigg \}ds\nonumber\\
&&+\frac{1}{4\pi\ap}\int _0^1\int _0^1\zb [f,A_N]\z (s)~\zb [f,A_M]\z (t)~
Y^N(s)Y^M(t)~q(s,t)dsdt~.\label{26a}
\eea
From this expression we can read off directly the vertices for a perturbative
evaluation of $\hat{\cal Z}$. Obviously, due to the triviality of our target space 
fields $(G,B,\Phi )$, there are no vertices in the bulk of the string world sheet, but on the  boundary only. The propagator $\langle Y^M(s)Y^N(t)\rangle $ restricted to the boundary is equal to $-2\ap \eta ^{MN}\log \vert
s-t\vert $. Since this is integrable, in lowest order the only divergences 
are due to the tadpole diagrams arising by connecting the $Y$-legs of
either the third or the fifth vertex in (\ref{26a}).
The divergence arising from the third term contains $\dot X^M$, hence it
contributes to the gauge field $\beta $-function. The fifth term contains
$\partial _{\bf n}X^1=\int \zb f\z q(s,t)dt$. This is an independent
divergence \cite{leigh,d} and constitutes the $\beta $-function for the 
brane position $f$. Altogether we find
\bea
\beta ^{(A)}_N&=&-\ap ~D^MF_{MN}\nonumber\\
\beta ^{(f)}&=&\frac{i}{2\pi}D^M[f,A_M]~=~\frac{1}{2\pi}D^MD_Mf~.
\label{27}
\eea
\section{$\beta $-functions in the case of dynamically realized \\Dirichlet boundary 
condition}
In this section we consider a model defined by a functional integral without
any constraint on the functional integrand and take as the action that
which arises by formulating the T-duality as a canonical transformation \cite{do}. Introducing
\beq
\breve W[X,\z ,\zb ]~=~\exp i\int _{\partial M}\left (\zb A_N\z ~\dot X^N
+\frac{1}{2\pi\ap }\zb f\z ~\partial _{\bf n}X_1\right )ds~,
\label{28}
\eeq
the partition function is given by
\bea
\lefteqn{\breve{\cal Z}[A,f]~=~\int DX^{\nu}~D\zb D\z ~\zb _a(0)\z _a(1)~e^{iS_0[\z , \zb]}}&&
\nonumber\\
&&\cdot ~\exp \left (\frac{i}{4\pi\ap}\int _Md^2z~\partial X^{\mu}\partial X_{\mu}-\frac{i}{2\pi\ap }\int _{\partial M}
X^1\partial _{\bf n}X_1ds\right )\breve W [X,\z ,\zb ]~.
\label{29}
\eea
The Dirichlet condition arises as part of the stationarity condition of the action
since $X^1-\zb f\z $ is the factor multiplying $\partial _{\bf n}\delta X_1$
on $\partial M$.
In contrast to the Wilson loops of the previous sections $\breve W$ is, due to
the presence of $\partial _{\bf n}X_1$, a functional of $X^{\mu}(z)$
not only on the boundary, but on the string world sheet $M$ itself.
Again, after a partial integration, we get as a starting point for the variational formulae
\bea
\frac{\delta\breve W}{\delta X^N(z)}&=&i~\breve W \int _0^1\delta ^{(2)}(z(s)-z)\label{31}\\
&\cdot &\left (\zb (\partial _NA_K-\partial _KA_N)\z \dot X^K+\frac{1}{2\pi\ap}
\zb \partial _Nf\z \partial _{\bf n}X_1-\dot{\zb }A_N\z -\zb A_N\dot{\z }
\right )ds~.
\nonumber
\eea
The total action for $\z ,\zb $ is
\bea
\breve S[\z ,\zb ]&=&S_0[\z ,\zb]-i\log \breve W\label{32}\\
&=&S_0[\z ,\zb]+\int _0^1\zb A_{N}\z \dot X^Nds+\frac{1}{2\pi\ap }\int _{\partial M}\zb f\z \partial _{\bf n}X_1ds~.\nonumber
\eea
With
\beq
\dot{\z} (t)~=~iA_{N}\z \dot X^{N}(t)+\frac{i}{2\pi\ap}f\z (t)\partial _
{\bf n}X_1 -~i \frac{\delta \breve S}{\delta \zb (s)}
\label{33}
\eeq
we repeat the procedure of section 2 and get
\beq
\frac{\delta \breve W}{\delta X^N(z)}~=~i\breve W\int ^1_0\delta ^{(2)}(z(s)-z)
\left ( \zb F_{NK}\z ~\dot X^K+\frac{1}{2\pi\ap }\zb D_Nf\z ~\partial _{\bf n}X_1
\right )ds~,\label{34}
\eeq
\bea
\lefteqn{\frac{\delta ^2\breve W}{\delta X^N(z)\delta X^M(z')}~=~-\breve W
\int ^1_0\delta ^{(2)}(z(s)-z)
\left ( \zb F_{NK}\z \dot X^K+\frac{1}{2\pi\ap }\zb D_Nf\z \partial _{\bf n}X_1
\right )ds}&&\nonumber\\
&&~~~~~~~~~~~~~~~~~~~~~~~~\cdot \int ^1_0\delta ^{(2)}(z(t)-z')
\left ( \zb F_{ML}\z \dot X^L+\frac{1}{2\pi\ap }\zb D_Mf\z \partial _{\bf n}X_1
\right )dt\nonumber\\
&&+i\breve W\int \delta ^{(2)}(z(s)-z')\delta ^{(2)}(z(s)-z)
\left (\zb D_MF_{NK}\z \dot X^K+\frac{1}{2\pi\ap }\zb D_MD_Nf\z \partial _{\bf n}X_1\right )ds\nonumber\\
&&+i\breve W\int \zb F_{NM}\z (s)~\delta ^{(2)}(z(s)-z)~\frac{d}{ds}\delta ^{(2)}(z(s)-z')ds~.
\label{35}
\eea
The variations with respect to $X^1$ are different qualitatively. There is
no way to get rid of $\partial _{\bf n}$ acting on the variation of $X^1$ by
some partial integration on the boundary. Thus no $\dot{\z}$ or $\dot{\zb }$
is generated, and we have no opportunity to apply the equations of motion
for $\z ,\zb $ in a useful manner.
\footnote{Fortunately here these manipulations are not needed to get manifest gauge invariant structures.}
The first and second derivatives are
\bea
\frac{\delta \breve W}{\delta X_1(z)}&=&\frac{i\breve W}{2\pi\ap}~\int
\zb f\z ~\partial _{\bf n}\delta ^{(2)}(z(s)-z)ds~,\label{36}\\
\frac{\delta ^2\breve W}{\delta X_1(z)\delta X_1(z')}&=&-\frac{\breve W}{(2\pi\ap )^2}~\int
\zb f\z ~\partial _{\bf n}\delta ^{(2)}(z(s)-z)ds~\int
\zb f\z ~\partial _{\bf n}\delta ^{(2)}(z(t)-z')dt~.
\nonumber
\eea
There is still a mixed derivative $\frac{\delta ^2\breve W}{\delta X^1(z)
\delta X^N(z')}$, but we save writing down its form and proceed to the
combination of all first and second derivatives in
\beq
\breve W[X+Y ,\z ,\zb ]=\breve W[X,\z ,\zb ]~\exp(i\breve{\Omega} [X,Y,\z ,\zb ])~,
\label{37}
\eeq
with
\bea
\breve{\Omega} [X,Y,\z ,\zb ]&=&
\int _0^1\bigg \{\Big (\zb F_{NL}\z (s)~
\dot X^{L}+\frac{1}{2\pi\ap}\zb D_Nf\z ~\partial _{\bf n}X_1\Big )Y ^{N}
 \nonumber\\
&&+\frac{1}{2\pi\ap} \zb f\z ~\partial _{\bf n}Y_1~+~\frac{1}{2}
 \zb F_{NM}\z ~\dot Y^MY^N\nonumber\\
&&+\frac{1}{2} \Big (\zb D_MF_{NK}\z ~\dot X^K+\frac{1}{2\pi\ap}\zb D_MD_Nf\z 
~\partial _{\bf n}X_1\Big )Y^MY^N\nonumber\\
&&+\frac{1}{4\pi\ap} \zb D_Mf\z ~Y^M \partial _{\bf n}Y_1
+O(Y^3)\bigg \}ds~.\label{38}  
\eea
To calculate the partition function (\ref{29}) we make a shift $X\rightarrow
X+Y$, use $D(X+Y)=DY$ and choose $X$ in such a way that the linear terms
in $Y$ vanish in the bulk of $M$, i.e. $\partial ^2 X^{\mu}=0$. However, we
do not specify any boundary condition on $X^{\mu}$.
\footnote{We could choose $X$ in such a way that all linear terms in $Y$
would vanish, but then the factorizing classical factor would contain
$\z ,\zb $ and mix with $\breve W[X,\z ,\zb ]$ and the $Y$ functional integral in doing the
$\z $-integration.}
Then we have to continue with
\bea
\breve{\cal Z}&=&\int D\zb D\z ~\zb _a(0)\z _a(1)~e^{iS_0[\z , \zb]+\frac{i}{4\pi\ap }\int (X^N\partial _{\bf n}X_N-X^1\partial _{\bf n}X_1)ds}~\breve W[X,\z ,\zb ]\label{39}\\
&&\cdot ~\int DY^{\nu}~\exp\left \{\frac{i}{4\pi\ap }\left (\int \partial Y^{\mu}
\partial Y_{\mu}d^2z-2\int Y^1\partial _{\bf n}Y_1ds\right )\right \}
\nonumber\\
&&~~~~~~~~~\cdot\exp\left \{\frac{i}{2\pi\ap}\int (Y_N\partial _{\bf n}X^N-X^1\partial _{\bf n}
Y_1)ds~+~i\breve{\Omega}[X,Y,\z ,\zb ]\right \}~.\nonumber
\eea
For the perturbative evaluation of the $Y$-integral it is convenient to
use
\bea
\lefteqn{\int DY^{\nu}~\exp\left \{\frac{i}{4\pi\ap}\int \partial Y^{\mu}\partial Y_{\mu}~
d^2z-\frac{i}{2\pi\ap}\int Y^1\partial _{\bf n}Y_1+i\int Y^{\mu}J_{\mu}
\right \}}&&\nonumber\\
&&=~(\mbox{det}\Delta )^{\frac{1}{2}}\exp \left \{\frac{-i}{2}\int J^{\alpha}(z)
\Delta _{\alpha\beta}(z,z')J^{\beta}(z')d^2zd^2z'\right \}~.
\label{40}
\eea
The propagator $\Delta$ of the above equation is defined by
\footnote{To avoid the discussion of the zero mode for the Neumann propagator
we take $M$ to be the upper half plane.}
\bea
\lefteqn{\Delta _{\mu\nu}~=~\eta _{\mu\nu}~\Delta _{\nu}~~~~~\mbox{(no sum)}~,}
&&\nonumber\\
&&\partial ^2\Delta _{\mu\nu}~=~-2\pi\ap \eta _{\mu\nu}\delta ^{(2)}(z-z')~,
~\mbox{in}M,~~~\Delta _1~=~0~,\partial _{\bf n}\Delta _N~=~0~,~\mbox{on}~\partial M~.
\label{41}
\eea 
Eqs. (\ref{39})-(\ref{41}) imply a set of Feynman rules with Neumann propagator
for $Y^N$, Dirichlet propagator for $Y^1$, no vertex in the bulk and boundary
vertices defined by $\breve{\Omega}$ as well as the two additional 1-leg
vertices in the last exponential of eq.(\ref{39}).

The two vertices in the third line of eq.(\ref{38}) yield divergent tadpole
contributions which would deliver just the same $\beta $-functions as in section 3, if they would be the only divergent diagrams. But in contrast to 
section 3, where 1-leg vertices appeared with underivated $Y$ only, now we
have the 1-leg vertex $\frac{i}{2\pi\ap}(\zb f\z -X^1)\partial _{\bf n}Y_1$.
The diagram connecting two such 1-leg vertices is
\beq
{\cal A}~\equiv \frac{1}{2}~\frac{-1}{(2\pi\ap )^2}\int dsdt~(\zb f\z (s)-X^1)
(\zb f \z (t)-X^1)~\partial _{\bf n}\partial _{\bf n}^{'}\Delta _1(z,z')\vert _{z=s,~z'=t}~.
\label{42}
\eeq
For the upper half plane
\beq
\Delta _1(z,z')~=~-\ap ~(\log \vert z-z'\vert ~-~\log \vert z-\bar{z'}\vert )
\label{43}
\eeq
leads to
\beq
{\cal A}~=~\frac{-1}{4\pi ^2\ap}\int dsdt~\frac{(\zb f\z (s)-X^1)(\zb f\z (t)-
X^1)}{(s-t)^2}~.
\label{44}
\eeq
${\cal A}$ is linearly divergent. We neglect linear divergences
\footnote{They are irrelevant in dimensional regularisation.} and look
only for logarithmic ones which could be produced by expanding the nominator
in  ${\cal A}$. If this nominator would be continuously differentiable, due to the antisymmetry of $(s-t)^{-1}$, there would be no such divergence. However,
we have to evaluate the diagram with quantised fields $Y$ $and$ $\z ,\zb $.
As a consequence the left and right limits of the derivative of the
nominator could differ. Anticipating this possibility we expand the nominator
for $s>t$ and $s<t$ separately
\bea
\zb f\z (s)~\zb f\z (t)&=&O(1)~+~(s-t)\bigg \{
\Theta (s-t)~\frac{d}{dt}\left (\zb f\z (t+0)\right )\zb f\z (t)\nonumber\\
&+& \Theta (t-s)~\frac{d}{dt}\left (\zb f\z (t-0)\right )\zb f\z (t)
\bigg \}~+~O((s-t)^2)~.
\label{45} 
\eea
Using the equations of motion (\ref{33}) this turns into
\bea
\zb f\z (s)~\zb f\z (t)&=&O(1)~+~(s-t)~\dot X^N~\bigg \{
\Theta (s-t)~\zb D_Nf\z (t+0)~\zb f\z (t)\nonumber\\
&+& \Theta (t-s)~\zb D_Nf\z (t-0)~\zb f\z (t)\bigg \}~+~O((s-t)^2)~.
\label{46} 
\eea
Now from the special structure of the $\z $-propagator it is obvious
that for the purpose of the perturbative evaluation of the $\z ,\zb $ functional
integral we can make the following identifications
\bea
\zb _a(t+0)~\z _b(t+0)~\zb _c(t)~\z _d(t)~=~\delta _{bc}~\zb _a(t+0)~\z _d(t)~,
\nonumber\\
\zb _a(t-0)~\z _b(t-0)~\zb _c(t)~\z _d(t)~=~\delta _{ad}~\zb _c(t)~\z _b(t-0)
~.\label{47} 
\eea
Using this in (\ref{46}) we get for the regularised version ($a$ regul. parameter) of (\ref{44})
\footnote{The terms in the nominator of (\ref{44}) containing f in zero or first order are continuously differentiable.} 
\bea
{\cal A}&=&\frac{-1}{4\pi ^2\ap}\int ~\bigg (\zb (D_Nf)f\z (t)\int _{s>t}\frac{s-t}{(s-t)^2+a^2}~ds\nonumber\\
&&~~~~~~~~~~~~~~~~+\zb fD_Nf\z (t)\int _{s<t}\frac{s-t}{(s-t)^2+a^2}~ds\bigg )
\dot X^N(t)dt~+~...~~,
\nonumber
\eea
the dots standing for either linear divergent or finite contributions.
This implies
\beq
{\cal A}~=~\frac{-1}{4\pi ^2\ap}\log a~\int ~\dot X^N\zb ~[f,D_Nf]~\z (t)dt~+~...~~.
\label{48}
\eeq 
This divergence is responsible for the presence of the mass term in the gauge field $\beta $-function. Altogether we find
\beq
\beta ^{(A)}_N~=~-\ap ~D^M F_{MN}~+~\frac{i}{4\pi ^2\ap}[f,D_N f]~.
\label{49}
\eeq
$\beta ^{(f)}$ coincides with that of section 3.

\section{Conclusions}
At least up to the considered order of perturbation theory, our results in 
section 3 and 4 clearly favour the D-brane model with 
dynamically implemented Dirichlet boundary conditions as the correct realization 
of the T-dual partner of the open string with free ends. The model with 
externally realized boundary conditions has been motivated \cite{do,d} by 
formal manipulations of the functional integral. Its failure to produce 
the mass term can be explained just by this formal nature. However, on a more 
ambitious level we would like to localise the forbidden step in those 
manipulations. A first idea we get by looking back to the application 
of the naive T-duality rules to the gauge field $\beta $-function of the 
open string with free ends. The mass term arises from taking the value $1$ 
for the summation index. The underlying ultraviolet divergence is due to 
a propagator in $1$-direction at coinciding points. Since there is no 
propagation in this direction in the model of section 3, it should be 
not surprising that no mass term appears. On the more technical level 
the comparison of the details of the calculations is useful. 
 
In both cases we consider the partition functions $\hat{\cal Z},~\breve{\cal Z}$ 
as defined by the interacting quantum field theory of the string world 
sheet position $X$ and the auxiliary fields $\z ,\zb $ in the sense of a 
combined functional integral. This point of view is implemented by the 
use of the $\z $ equations of motion including contact terms for the expansion 
of the modified Wilson loops. 
\footnote{Note that in $both$ cases the $\z $ equations of motion are compatible 
with naive duality.} 
However, we have given up this understanding at one occasion, namely 
when we switched to a partially stepwise evaluation by performing the $X^1$ 
integration in section 3. But just the intermediate treatment of $\z ,\zb $ 
as classical fields is the dangerous step as can be seen in the calculation 
of diagram ${\cal A}$ in section 4. If we there had treated $\z ,\zb $ as 
classical fields we would also have lost the mass term. 
 
The stepwise, i.e. first $X$ and then $\z ,\zb $ integration has been 
crucial for giving the formal notion of matrix valued brane position a 
technical meaning via insertion of matrices in the argument which 
specifies an explicit boundary parameter dependent scalar valued Dirichlet 
boundary condition \cite{do,d}. Now we see that just this stepwise procedure 
violates the equivalence with the original free end model. Nevertheless 
the model of \cite{d} is well defined and interesting for its own.

We stress that the preference of the dynamically over the external realization 
of boundary conditions developed in this paper concerns the case of 
non-Abelian gauge fields and the implementation of path ordering via 
the $\z $-formalism only. In the Abelian case we see no stumbling block 
in the equivalent transformation of the free end model to the D-brane model 
with externally realized boundary conditions. We can also not exclude that 
in the non-Abelian case there could exist T-dual versions with external 
boundary conditions beyond the $\z $-formalism, for some alternative aspects 
of the construction of a $\sigma $-model describing matrix D-branes 
see \cite{mav}.

Our discussion concerned the test of quantum equivalence of naive T-dual
models in the general, not necessary conformally invariant situation. Since 
T-duality as a map between equivalent theories is by far better established
in the conformal case, it would be interesting to understand both the 
potential differences between externally and dynamically realized boundary
conditions as well as the $\sigma$-model description of boundaries coupling
to non-Abelian gauge fields in the formalism of boundary conformal field
theories, \cite{refu} and refs. therein.  
\\[5mm]
{\bf Acknowledgement:} I would like to thank H.-J. Otto, C. Preitschopf,
A. Recknagel and V. Schomerus for useful discussions.

\end{document}